\documentclass[aps,prd,twocolumn,groupedaddress,showpacs]{revtex4}
\usepackage{amsfonts,amssymb,amsmath,mathrsfs}
\begin{document}
\title{$AdS_4\times\mathbb{CP}^3$ superstring and $D=3$ $\mathcal N=6$ superconformal symmetry}
\author{D.V.~Uvarov}
\email{d_uvarov@hotmail.com; uvarov@kipt.kharkov.ua}
\affiliation{NSC Kharkov Institute of Physics and Technology,\\
61108 Kharkov, Ukraine}
\date{\today}

\begin{abstract}
Motivated by the isomorphism between $osp(4|6)$ superalgebra and
$D=3$ $\mathcal N=6$ superconformal algebra we consider the
superstring action on the $AdS_4\times\mathbb{CP}^3$ background
parameterized by $D=3$ $\mathcal N=6$ super-Poincare and
$\mathbb{CP}^3$ coordinates  supplemented by the coordinates
corresponding to dilatation and superconformal generators. It is
also discussed the relation between the degeneracy of fermionic
equations of motion and the action $\kappa-$invariance in the
framework of the supercoset approach.
\end{abstract}
\pacs{11.25.-w, 11.30.Pb, 11.25.Tq}
\maketitle

\section{\label{sec1}Introduction}

The idea of gauge/string correspondence has been elaborated since
the early days of string theory. During the last decade
significant progress has been attained in understanding the
duality \cite{Maldacena}-\cite{Witten:1998qj} between the $D=4$ $\mathcal N=4$ superYang-Mills theory and
string theory on $AdS_5\times S^5$ background. Recently novel
example of the gauge/string correspondence has been proposed \cite{ABJM} involving the
superconformal $D=3$ $\mathcal N=6$ Chern-Simons-matter theory \footnote{This $3d$ superconformal field theory describes the low energy limit of the world-volume theory on $N$ M2-branes probing the $\mathbf C^4/\mathbb Z_k$ singularity. Significant interest to the exploration of $3d$ superconformal field theories in the context of M2-brane low energy world-volume dynamics has been provoked by \cite{BLG}.} with the
gauge group $U(N)\times U(N)$ and level $k$ and M-theory on
$AdS_4\times(S^7/\mathbb Z_k)$ background. In the t'Hooft
limit $N,k\to\infty$ with $\lambda=N/k$ fixed the field theory can be effectively described by the $IIA$ superstring
on $AdS_4\times\mathbb{CP}^3$ background.

For both dualities one of the main unsolved problems is to
quantize corresponding superstring models. The full action for the
Green-Schwarz (GS) superstring on $AdS_5\times S^5$ was constructed in
\cite{MT}, \cite{Kallosh} on the symmetry grounds using that
$AdS_5\times S^5$ is the maximally supersymmetric background of
Type $IIB$ supergravity and that all bosonic and fermionic degrees
of freedom fit into the supercoset space
$PSU(2,2|4)/(SO(1,4)\times SO(5))$. It was then discovered
\cite{BPR} that such full action is classically integrable
extending the previous result \cite{Wadia} for the bosonic model.
This stimulated application of the methods developed for the
investigation of integrable systems \footnote{For recent review see
e.g. \cite{Dorey} and references therein.}. However, the
nonlinearity of the superstring action even after the exclusion of the pure
gauge degrees of freedom still precludes from solving the
quantization problem and motivates application of the approximate
methods \cite{FT}, \footnote{In analogy with the $AdS_5\times S^5$
case the semiclassical quantization around the particular solutions of
string equations on $AdS_4\times\mathbb{CP}^3$ has been considered
in \cite{semiclas}.}.

To obtain the superstring action on $AdS_4\times\mathbb{CP}^3$
background including the fermions it has been suggested in \cite{AF}, \footnote{See also
\cite{Stefanski} and \cite{PS} for the discussion of $AdS_4\times\mathbb{CP}^3$ superstring in
the pure spinor formulation.} to apply the
supercoset method of \cite{MT}. The main observation
is that the bosonic degrees of freedom fit into the bosonic body
$(Sp(4)/SO(1,3))\times(SO(6)/U(3))$ of the supercoset space
$OSp(4|6)/(SO(1,3)\times U(3))$ that also allows to accommodate 24
fermions equal in number to the supersymmetries preserved by the $AdS_4\times\mathbb{CP}^3$
background. It was shown \cite{AF} that such superstring
action involving 24 fermions is invariant under the $8-$parameter
$\kappa-$symmetry transformations and is classically integrable.

Similarly to the $AdS_5\times S^5$ superstring the original
superstring action on $AdS_4\times\mathbb{CP}^3$ was given in the
$AdS$ basis for the Cartan forms with the appropriate choice of
the supercoset element. The isomorphism between the $AdS_4$
algebra and conformal algebra in 1+2 dimensions suggests considering also
the superstring action in the conformal basis \footnote{The $AdS_5\times
S^5$ superstring action in the conformal basis was studied in
\cite{MTlc}, \cite{Metsaev}.}. Choosing the $OSp(4|6)/(SO(1,3)\times U(3))$ supercoset representative
parametrized by the $D=3$ $\mathcal N=6$ superspace coordinates,
$\mathbb{CP}^3$ coordinates and those associated with the dilatation
and superconformal generators yields the action with manifest
$D=3$ $\mathcal N=6$ super-Poincare symmetry that is the subgroup
of the symmetry group on the field theory side of the duality
\cite{ABJM}, \cite{BLS}.

It should be noted that despite the fact that the supercoset action on the
$AdS_4\times\mathbb{CP}^3$ background has clear group-theoretical
structure, involves the correct number of physical degrees of freedom
and is classically integrable, unlike the supercoset action on
$AdS_5\times S^5$, it cannot describe all possible superstring
motions, as was already observed in \cite{AF}. To study such
string configurations the explicit form of the action depending on
all 32 fermionic variables is needed that in turn requires to
elaborate on the full superspace solution of the $IIA$ supergravity on
$AdS_4\times\mathbb{CP}^3$ \cite{GSWnew}. However, whether such
full-fledged action for the $IIA$ superstring on
$AdS_4\times\mathbb{CP}^3$ inherits the integrability property remains
unknown.

In section \ref{sec2} we discuss the properties of the superstring action
on the supercoset space $OSp(4|6)/(SO(1,3)\times U(3))$. In particular, the equations of motion for the
fermions are cast into the form close to that derived in the
conventional GS approach \cite{GS} and it is
proved that in the general case 8 of 24 equations are trivial. So
that the $\kappa-$symmetry already manifests itself at the level of the
equations of motion. This degeneracy of the equations of motion for
the fermions traces back to the form of the anticommutators of the
fermionic generators of $D=3$ $\mathcal N=6$ superconformal
algebra. We also give the representation for the $\kappa-$symmetry
transformations, that allow to gauge away $\frac13$ of the
fermionic degrees of freedom, in the form amenable for comparison
with the GS $\kappa-$symmetry transformations that remove
$\frac12$ of the fermions.

In section \ref{sec3} we derive the explicit expressions for the Cartan
forms in the conformal basis starting from the $OSp(4|6)/(SO(1,3)\times U(3))$
supercoset element and use them to write the superstring action in the form with manifest $D=3$ $\mathcal N=6$ super-Poincare symmetry. It is discussed the
possibility of fixing the gauge freedom related to the 8-parameter $\kappa-$symmetry.

In the appendixes there are summarized the relevant properties of
spinors and $\gamma-$matrices in $D=2+3$, $D=1+3$ and $D=1+2$
dimensions, and there are given the details on the isomorphism
between the $osp(4|6)$ superalgebra and $D=3$ $\mathcal N=6$
superconformal algebra.

\section{\label{sec2}Superstring action in the supercoset approach: equations of motion and $\kappa-$symmetry}

The starting point is the $OSp(4|6)/(SO(1,3)\times U(3))$ supercoset element $\mathscr G$ that is used to
define the left-invariant Cartan 1-forms
\begin{eqnarray}\label{defcartan}
\mathscr G^{-1}d\mathscr
G&=&G_{\underline{mn}}(d)M^{\underline{mn}}+\Omega_{\hat
i}{}^{\hat j}(d)V_{\hat j}{}^{\hat i}\nonumber\\
&+& F^\alpha_a(d)O^a_\alpha+\bar F^{\alpha a}(d)\bar O_{\alpha
a}.\nonumber\\
\end{eqnarray}
The bosonic 1-forms $G_{\underline{mn}}(d)$,
$\underline{m},\underline{n}=0',0,...,3$ are associated with the
$so(2,3)\sim sp(4)$ generators $M^{\underline{mn}}$ that can be
split into the $so(1,3)$ generators $M^{m'n'}$, $m',n'=0,...,3$ and the
$so(2,3)/so(1,3)$ coset generators $M^{0'm'}$ that corresponds to
representing the $so(2,3)$ algebra as the $AdS_4$ one. Accordingly the 1-forms $G_{m'n'}(d)$ define the $so(1,3)$ connection and $G_{0'm'}(d)$ the $AdS_4$ veirbein.
Analogously the
Cartan forms $\Omega_{\hat i}{}^{\hat j}(d)$,
$\hat i,\hat j=1,...,4$
\begin{equation}\label{su4cf}
\Omega_{\hat i}{}^{\hat j}=\left(
\begin{array}{cc}
\Omega_a{}^b& \Omega_a{}^4\\
\Omega_4{}^b& \Omega_4{}^4
\end{array}\right),\qquad \Omega_4{}^4=-\Omega_a{}^a
\end{equation}
can be split into the 1-forms $\Omega_a{}^b(d)$ corresponding to the $u(3)$ generators $V_a{}^b$ and the 1-forms $\Omega_4{}^a(d)$, $\Omega_a{}^4(d)$ related to the
$su(4)/u(3)$ coset generators $V_a{}^4$, $V_4{}^a$. These forms define the $u(3)$ connection and the $\mathbb{CP}^3$ vielbein, respectively.
The fermionic 1-forms $F^\alpha_a(d)$ and $\bar F^{\alpha
a}(d)$ are related to the $osp(4|6)$ odd generators
$O^a_\alpha$, $\bar O_{\alpha a}$ carrying the $D=2+3$
Majorana spinor index $\alpha=1,...,4$ and transforming in the vector representation of $SO(6)$ that decomposes as $\mathbf3\oplus\bar{\mathbf3}$
with respect to $SU(3)$ (see Appendix B). By construction the Cartan forms (\ref{defcartan}) satisfy the Maurer-Cartan (MC)
equations that can be schematically written as
\begin{equation}
d\omega_{\mathcal A}+\frac12\omega_{\mathcal B}(d)\wedge\omega_{\mathcal C}(d)f^{\mathcal{CB}}{}_{\mathcal A}=0,
\end{equation}
where $f^{\mathcal{CB}}{}_{\mathcal A}$ are the structure constants of the $osp(4|6)$ superalgebra.

Under the discrete automorphism $\Upsilon$ of the $osp(4|6)$ superalgebra the $so(1,3)$
and $u(3)$ generators are inert, while the remaining bosonic
generators change the sign $\Upsilon(M^{0'm'})=-M^{0'm'}$,
$\Upsilon(V_a{}^4)=-V_a{}^4$, $\Upsilon(V_4{}^a)=-V_4{}^a$.
The fermionic generators can be split into 2 eigenspaces $P_{\pm\alpha}\vphantom{P}^\beta O^a_\beta$: 
$\Upsilon(P_{\pm\alpha}\vphantom{P}^\beta O^a_\beta)=\pm iP_{\pm\alpha}\vphantom{P}^\beta O^a_\beta$ using the $4d$ chiral projectors introduced in Eq.(\ref{chiralp}). These transformations of the $osp(4|6)$
generators induce transformations of the associated Cartan forms and
serve as the guide to construct the $\mathbb Z_4$-invariant
superstring action
\begin{equation}\label{action}
\mathscr S=-\frac12\int
d^2\xi\sqrt{-g}g^{ij}(G_{i0'}{}^{m'}G_{j0'm'}+\Omega_{ia}{}^4\Omega_{j4}{}^a)+\mathscr
S_{WZ},
\end{equation}
where the Wess-Zumino term is given by the wedge product of the fermionic Cartan forms \cite{Berkovits}, \cite{RS}
\begin{equation}
\mathscr S_{WZ}=\frac{i}{2}\varepsilon^{ij}\int d^2\xi
F^\alpha_{ia}C'_{\alpha\beta}\bar F_j^{\beta a}.
\end{equation}
Two summands entering the kinetic term correspond to the $AdS_4$ and
$\mathbb{CP}^3$ parts of the background. The WZ term involves the
$D=1+3$ charge conjugation matrix $C'_{\alpha\beta}$.

The superstring Lagrangian is constructed out of the world-sheet projections of Cartan 1-forms; thus to find its variation it is necessary to
consider the variations of relevant 1-forms. Using the general formula for the variation of a form
\begin{equation}
\delta F(d)=d(i_\delta F(d))+i_\delta(dF(d))
\end{equation}
in the second summand one substitutes the MC equations
\begin{eqnarray}
dG^{0'm'}&-&2G^{m'}{}_{n'}(d)\wedge
G^{0'n'}(d)\nonumber\\
&-&iF^\alpha_a(d)\wedge\gamma^{0'm'}_{\alpha\beta}\bar
F^{\beta a}(d)=0,\nonumber \\
\end{eqnarray}
\begin{eqnarray}
d\Omega_a{}^4&+&i\Omega_{+a}{}^b(d)\wedge\Omega_b{}^4(d)\nonumber\\
&-&\varepsilon_{abc}\bar
F^{\alpha b}(d)\wedge C_{\alpha\beta}\bar F^{\beta
c}(d)=0,\nonumber \\
\end{eqnarray}
\begin{eqnarray}
d\Omega_4{}^a&+&i\Omega_4{}^b(d)\wedge\Omega_{+b}{}^a(d)\nonumber\\
&+&\varepsilon^{abc}F^{\alpha}_{b}(d)\wedge
C_{\alpha\beta}F^{\beta}_{c}(d)=0,\nonumber \\
\end{eqnarray}
\begin{eqnarray}
dF^\alpha_a&+&\frac12F^\beta_a(d)\wedge
G_{\underline{mn}}(d)\gamma^{\underline{mn}}{}_\beta{}^\alpha
+i\Omega_{-a}{}^b(d)\wedge F^\alpha_b(d)\nonumber\\
&+&i\varepsilon_{acb}\Omega_4{}^c(d)\wedge\bar F^{\alpha
b}(d)=0,\nonumber \\
\end{eqnarray}
\begin{eqnarray}
d\bar F^{\alpha a}&+&\frac12\bar F^{\beta a}(d)\wedge
G_{\underline{mn}}(d)\gamma^{\underline{mn}}{}_\beta{}^\alpha+i\bar
F^{\alpha b}(d)\wedge\Omega_{-b}{}^a(d)\nonumber\\
&-&i\varepsilon^{acb}\Omega_c{}^4(d)\wedge F^{\alpha}_{b}(d)=0,\nonumber \\
\end{eqnarray}
where
$G_{\underline{mn}}(d)\gamma^{\underline{mn}}{}_\alpha{}^\beta=2G_{0'm'}(d)\gamma^{0'm'}{}_\alpha{}^\beta
+G_{m'n'}(d)\gamma^{m'n'}{}_\alpha{}^\beta$ and
$\Omega_{\pm a}{}^b(d)=\Omega_a{}^b(d)\pm\delta_a^b\Omega_c{}^c(d)$
are the $so(2,3)$ and $u(3)\oplus u(1)$ connections.
Then the variation of the 1-forms entering the action (\ref{action}) acquires the form
\begin{eqnarray}
\delta G^{0'm'}(d)&=&dG^{0'm'}(\delta)\nonumber\\
&+&2G^{m'}{}_{n'}(d)G^{0'n'}(\delta)-2G^{m'}{}_{n'}(\delta)G^{0'n'}(d)\nonumber\\
&+&iF^\alpha_a(d)\gamma^{0'm'}_{\alpha\beta}\bar F^{\beta
a}(\delta)-iF^\alpha_a(\delta)\gamma^{0'm'}_{\alpha\beta} \bar
F^{\beta a}(d),\nonumber\\
\end{eqnarray}
\begin{eqnarray}
\delta\Omega_a{}^4(d)&=&d\Omega_a{}^4(\delta)-i\Omega_{+a}{}^b(d)\Omega_b{}^4(\delta)+i\Omega_{+a}{}^b(\delta)
\Omega_b{}^4(d)\nonumber
\\
&+&2\varepsilon_{abc}\bar F^{\alpha b}(d)C_{\alpha\beta}\bar
F^{\beta c}(\delta),\nonumber\\
\end{eqnarray}
\begin{eqnarray}
\delta\Omega_4{}^a(d)&=&d\Omega_4{}^a(\delta)-i\Omega_4{}^b(d)\Omega_{+b}{}^a(\delta)+i\Omega_4{}^b(\delta)\Omega_{+b}{}^a(d)
\nonumber\\
&-&2\varepsilon^{abc}F^{\alpha}_{b}(d)C_{\alpha\beta}F^{\beta}_{c}(\delta),\nonumber\\
\end{eqnarray}
\begin{eqnarray}
\delta F^\alpha_a(d)&=&dF^\alpha_a(\delta) -\frac12
F^\beta_a(d)G_{\underline{mn}}(\delta)\gamma^{\underline{mn}}{}_\beta{}^\alpha\nonumber\\
&+&\frac12 F^\beta_a(\delta)G_{\underline{mn}}(d)\gamma^{\underline{mn}}{}_\beta{}^\alpha\nonumber\\
&-&i\Omega_{-a}{}^b(d)F^\alpha_b(\delta)+i\Omega_{-a}{}^b(\delta)F^\alpha_b(d)
\nonumber\\
&-&i\varepsilon_{acb}\Omega_4{}^c(d)\bar F^{\alpha
b}(\delta)+i\varepsilon_{acb}\Omega_4{}^c(\delta)\bar F^{\alpha
b}(d),\nonumber\\
\end{eqnarray}
\begin{eqnarray}
\delta\bar F^{\alpha a}(d)&=&d\bar F^{\alpha
a}(\delta)-\frac12\bar F^{\beta a}(d)
G_{\underline{mn}}(\delta)\gamma^{\underline{mn}}{}_\beta{}^\alpha\nonumber\\
&+&\frac12\bar F^{\beta a}(\delta)
G_{\underline{mn}}(d)\gamma^{\underline{mn}}{}_\beta{}^\alpha\nonumber \\
&-& i\bar F^{\alpha b}(d)\Omega_{-b}{}^a(\delta)+i\bar F^{\alpha
b}(\delta)\Omega_{-b}{}^a(d)\nonumber \\
&+&i\varepsilon^{acb}\Omega_c{}^4(d)F^{\alpha}_{b}(\delta)-i\varepsilon^{acb}\Omega_c{}^4(\delta)F^{\alpha}_{b}(d).\nonumber\\
\end{eqnarray}

Since of the utmost importance is the $\kappa-$invariance of the
superstring action (\ref{action}) we concentrate on the
fermionic contribution to the variation of the action
\begin{eqnarray}\label{fermivar}
\delta\mathscr S|_{f}&=&\int d^2\xi\left(\bar{\mathscr F}^{\alpha\hat
a}_{\{+\}i}V^{ij}_+M_{j\alpha\hat a}{}^{\beta\hat
b}C_{\beta\gamma}\mathscr F^\gamma_{\{+\}\hat b}(\delta)\right.\nonumber\\
&+&
\left.\bar{\mathscr F}^{\alpha\hat a}_{\{-\}i}V^{ij}_-M_{j\alpha\hat
a}{}^{\beta\hat b}C_{\beta\gamma}\mathscr F^\gamma_{\{-\}\hat
b}(\delta)\right),\nonumber\\
\end{eqnarray}
where
\begin{equation}
M_{i\alpha\hat a}{}^{\beta\hat
b}=\left(\begin{array}{cc}
-i\delta^b_aG_{i0'm'}\gamma^{0'm'}{}_{\alpha}{}^\beta & \delta_\alpha^\beta\varepsilon_{acb}\Omega_{i4}{}^c\\
-\delta_\alpha^\beta\varepsilon^{acb}\Omega_{ic}{}^4 &
-i\delta^a_bG_{i0'm'}\gamma^{0'm'}{}_{\alpha}{}^\beta
\end{array}\right).
\end{equation}
The expression (\ref{fermivar}) analogously to the GS superstring case \cite{GS} involves the world-sheet projectors
\begin{equation}\label{wsp}
V^{ij}_{\pm}=\frac12(\sqrt{-g}g^{ij}\pm\varepsilon^{ij})
\end{equation}
obeying the relations
\begin{eqnarray}
V^{ij}_++V^{ij}_-=\sqrt{-g}g^{ij},&&
V^{ik}_{\pm}g_{kl}V^{jl}_{\pm}=0,\nonumber\\
V^{ik}_{\pm}g_{kl}V^{lj}_{\pm}=\sqrt{-g}V^{ij}_{\pm},&&
V^{ij}_{\pm}V^{kl}_{\pm}=V^{kj}_{\pm}V^{il}_{\pm}.\nonumber\\
\end{eqnarray}
The fermionic variation parameters and the world-sheet projections of Cartan forms have been grouped as follows:
\begin{equation}
\mathscr F^\alpha_{\{\pm\}\hat a}(\delta)={F^\alpha_{(\pm)a}(\delta)\choose\bar F^{\alpha a}_{(\mp)}(\delta)}
\end{equation}
and
\begin{equation}
\bar{\mathscr F}^{\alpha\hat a}_{\{\pm\}i}={\bar F^{\alpha a}_{(\mp)i}\choose F^\alpha_{(\pm)ia}}.
\end{equation}
They include chiral in the $D=1+3$ dimensional sense spinors
$F^\alpha_{(\pm)a}(\delta)$, $F^\alpha_{(\pm)ia}$ and their
conjugates $\bar F^{\alpha a}_{(\mp)}(\delta)$, $\bar F^{\alpha
a}_{(\mp)i}$ \footnote{The $4d$ chirality of the fermionic variation
parameters $F^\alpha_{(\pm)a}(\delta)$, $\bar F^{\alpha
a}_{(\mp)}(\delta)$ and Cartan 1-forms $F^\alpha_{(\pm)a}(d)$,
$\bar F^{\alpha a}_{(\mp)}(d)$ is labeled by the lower sign index
in parantheses. Note that in $D=1+3$ dimensions complex conjugation reverses
the chirality. Spinors $\mathscr F^\alpha_{\{\pm\}\hat a}(\delta)$ and
$\bar{\mathscr F}^{\alpha\hat a}_{\{\pm\}i}$ are nonchiral, however,
to distinguish from one another we have still endowed them with
the chirality labels in braces of the corresponding
unbarred components.}. The chiral projectors are defined as
\begin{eqnarray}\label{chiralp}
P_+^\alpha{}_\beta&=&\frac12(\delta^\alpha_\beta+C^{\alpha\gamma}C'_{\gamma\beta}),\nonumber\\
P_-^\alpha{}_\beta&=&(P_+^\alpha{}_\beta)^*=
\frac12(\delta^\alpha_\beta-C^{\alpha\gamma}C'_{\gamma\beta})\nonumber\\
\end{eqnarray}
and satisfy the requisite properties
\begin{eqnarray}
&P_++P_-=I,&\nonumber\\
&P_{\pm}P_{\pm}=P_{\pm},\qquad P_+P_-=P_-P_+=0,&\nonumber\\
\end{eqnarray}
because of the relation
\begin{equation}
C^{\alpha\beta}C'_{\beta\gamma}C^{\gamma\delta}C'_{\delta\epsilon}=\delta^\alpha_\epsilon.
\end{equation}
The definition (\ref{chiralp}) is justified by the fact that $C^{\alpha\beta}C'_{\beta\gamma}$ is related to the $4d$ matrix
$\Gamma^5_\alpha{}^\beta$
\begin{equation}
C^{\alpha\beta}C'_{\beta\delta}=iC'^{\alpha\beta}\Gamma^5_\beta{}^\gamma
C'_{\gamma\delta}.
\end{equation}
To derive (\ref{fermivar}) we have also used the following properties of chiral projectors (\ref{chiralp})
\begin{equation}
P_{\pm}^\alpha{}_\beta
C_{\alpha\gamma}P_{\mp}^\gamma{}_\delta=0,\qquad
P_{\pm}^\alpha{}_\beta\gamma^{0'm'}_{\alpha\gamma}P_{\pm}^\gamma{}_\delta=0.
\end{equation}

The variation (\ref{fermivar}) determines the equations of motion for the fermions
\begin{equation}\label{fermieom}
V^{ij}_{\pm}M^T_j{}^{\alpha\hat a}{}_{\beta\hat b}\bar{\mathscr
F}^{\beta\hat b}_{\{\pm\}i}=0.
\end{equation}
To find whether all of the equations (\ref{fermieom}) are nontrivial we need to compute the rank of $M^T_j{}^{\alpha\hat a}{}_{\beta\hat b}$
\begin{equation}
M^T_j{}^{\alpha\hat a}{}_{\beta\hat b}=\left(\begin{array}{cc}
-i\delta^a_bG_{j0'm'}\gamma^{0'm'}{}^{\alpha}{}_\beta & \delta^\alpha_\beta\varepsilon^{acb}\Omega_{jc}{}^4\\
-\delta^\alpha_\beta\varepsilon_{acb}\Omega_{j4}{}^c & -i\delta^b_aG_{j0'm'}\gamma^{0'm'}{}^{\alpha}{}_\beta
\end{array}\right)
\end{equation}
on shell of the Virasoro constraints
\begin{eqnarray}\label{vir}
\frac{\delta\mathscr S}{\delta
g^{ij}(\xi)}&=&G_{i0'm'}G_{j0'}{}^{m'}+\frac12(\Omega_{ia}{}^4\Omega_{j4}{}^a+\Omega_{ja}{}^4\Omega_{i4}{}^a)\nonumber\\
&-&\frac12g_{ij}g^{kl}(G_{k0'm'}G_{l0'}{}^{m'}+\Omega_{ka}{}^4\Omega_{l4}{}^a)=0.\nonumber\\
\end{eqnarray}
In equations (\ref{fermieom}) the $2d$ vector index of $M^T_j{}^{\alpha\hat a}{}_{\beta\hat b}$ is acted by the world-sheet projectors (\ref{wsp})
so that only one of its components out of two is independent. This can be illustrated, for instance, by the action of
$V^{ij}_{\pm}$ on a vector $F_j$ that can be presented as
\begin{eqnarray}
&V^{ij}_{\pm}F_j=V^i_{\pm}F_{\pm\tau},\qquad
V^i_{\pm}=\frac12\left(\begin{array}{c}
1\\
\frac{\sqrt{-g}g^{\tau\sigma}\mp1}{\sqrt{-g}g^{\tau\tau}}
\end{array}\right),&\nonumber\\
&F_{\pm\tau}=\sqrt{-g}g^{\tau\tau}F_\tau+(\sqrt{-g}g^{\tau\sigma}\pm1)F_\sigma.&\nonumber\\
\end{eqnarray}
Similarly the result of $V^{ij}_{\pm}$ projector action on the Virasoro constraints (\ref{vir}) reads
\begin{eqnarray}\label{vir'}
V^{ik}_{\pm}V^{jl}_{\pm}\frac{\delta\mathscr S}{\delta
g^{kl}(\xi)}&=&V^i_{\pm}V^j_{\pm}(G_{\pm\tau0'm'}G_{\pm\tau0'}{}^{m'}\nonumber\\
&+&\Omega_{\pm\tau
a}{}^4\Omega_{\pm\tau4}{}^a)=0.\nonumber\\
\end{eqnarray}
Then one observes that the matrix $G_{\pm\tau0'm'}\gamma^{0'm'}{}^{\alpha}{}_\beta$ is nonsingular
\begin{equation}
G_{\pm\tau0'm'}\gamma^{0'm'}{}^{\alpha}{}_\beta
G_{\pm\tau0'n'}\gamma^{0'n'}{}^{\beta}{}_\gamma=G_{\pm\tau\pm\tau}\delta^\alpha_\gamma,
\end{equation}
where $G_{\pm\tau\pm\tau}=G_{\pm\tau0'm'}G_{\pm\tau0'}{}^{m'}$.
This allows by the rank preserving transformation to bring $M^T$
to the triangular form
\begin{equation}
\left(\begin{array}{cc}
-i\delta^a_bG_{\pm\tau0'm'}\gamma^{0'm'}{}^{\alpha}{}_\beta&
\delta^\alpha_\beta\varepsilon^{acb}\Omega_{\pm\tau c}{}^4\\
0& i\frac{\Omega_{\pm\tau
a}{}^4\Omega_{\pm\tau4}{}^b}{G_{\pm\tau\pm\tau}}G_{\pm\tau0'm'}\gamma^{0'm'}{}^{\alpha}{}_\beta\\
\end{array}\right).
\end{equation}
Since the rank of the $3\times3$ matrix $\Omega_{\pm\tau
a}{}^4\Omega_{\pm\tau4}{}^b$ is unity \footnote{This statement is similar to
that the $2\times2$ matrix
$p_{m'}\tilde\sigma^{m'\dot\alpha\alpha}$ corresponding to $D=1+3$
null vector $p_{m'}$ has unit rank. Then the Cartan-Penrose
representation $p_{m'}\tilde\sigma^{m'\dot\alpha\alpha}=\bar
u^{\dot\alpha}u^\alpha$ is used to solve the constraint
$p^{m'}p_{m'}=0$ \cite{Penrose}.}, the rank of $M^T$ equals
$4\times3+4\times1=16$. As a result 8 out of 24 equations
(\ref{fermieom}) are trivial and this implies via the second
Noether theorem the 8-parameter fermionic symmetry of the action
(\ref{action}). The crucial distinction of the supercoset
string model \cite{AF} from the GS superstring on flat background
\cite{GS} and on $AdS_5\times S^5$ \cite{MT} is that the
$\kappa-$symmetry can gauge away only $\frac13$ of the fermions rather
than $\frac12$. This is attributed to the fact that the action
(\ref{action}) could be obtained by the partial $\kappa-$symmetry
gauge fixing from the full action containing 32 fermionic degrees
of freedom and the 8-parameter fermionic symmetry of
(\ref{action}) is the remnant of the 16-parameter symmetry of that
full action \footnote{It has been observed in \cite{AF} that
whenever the string moves only in $AdS_4$ part of the background so
that $\Omega_{\pm\tau a}{}^4=\Omega_{\pm\tau4}{}^b=0$ the rank of the
fermionic equations of motion gets increased to 12 that formally
leads to the mismatch between the bosonic and fermionic degrees of
freedom. This means that to correctly describe such string motions
ones has to consider the full superstring action on
$AdS_4\times\mathbb{CP}^3$ rather than the supercoset sigma model. For detailed discussion on that point
see \cite{GSWnew}.}.

It is worthwhile to note that the matrix $M$ can be obtained starting from the matrix of the anticommutators of the fermionic generators of
the $osp(4|6)$ superalgebra
\begin{equation}
\left(\begin{array}{cc}
\{\bar O_{\alpha a},O^{b}_\beta\} & \{\bar O_{\alpha a},\bar O_{\beta b}\} \\
\{O^{a}_\alpha,O^{b}_\beta\} & \{O^{a}_\alpha,\bar O_{\beta b}\}
\end{array}\right).
\end{equation}
Substituting the explicit expressions for its entries (see Appendix B)
\begin{widetext}
\begin{equation}
\left(\begin{array}{cc}
-i\delta_a^b\gamma^{\underline{mn}}_{\alpha\beta}M_{\underline{mn}}+2C_{\alpha\beta}(V_a{}^b-\delta_a^bV_c{}^c)& 2C_{\alpha\beta}\varepsilon_{acb}V_4{}^c\\
-2C_{\alpha\beta}\varepsilon^{acb}V_c{}^4 &
-i\delta^a_b\gamma^{\underline{mn}}_{\alpha\beta}M_{\underline{mn}}-2C_{\alpha\beta}(V_b{}^a-\delta_b^aV_c{}^c)
\end{array}\right)
\end{equation}
\end{widetext}
and replacing the $so(2,3)/so(1,3)$ and $su(4)/u(3)$ coset generators by the Cartan forms
$M_{0'm'}\to G_{0'm'}(d)$, $V_a{}^4\to\Omega_a{}^4(d)$, $V_4{}^a\to\Omega_4{}^a(d)$ yields up to the overall factor the entries
of the matrix $M$. This is of course the anticipated result since the action variation is determined by the variation of Cartan
forms that in turn depends on the structure constants of the $osp(4|6)$ superalgebra. However, this observation could be
of more use when applied backwards: starting from the matrix composed of the anticommutators of the fermionic generators of the isometry
superalgebra for some superbackground, whose bosonic part can be presented as the coset space,
one can study the degeneracy of such a matrix to find whether the corresponding string model,
constructed using the supercoset approach, will be $\kappa-$invariant.

Let us consider the $\kappa-$invariance property of the action (\ref{action}) in more detail. An equal number of physical and pure gauge fermions in the GS
superstring implied that the same matrix with the space-time spinor indices was present both in the
equations of motion for the fermions and the $\kappa-$symmetry
transformation rules. However, in the present case $M^T$ can not directly
appear in the $\kappa-$transformations because it is required that the
matrix of the rank 8 single out the requisite number of independent
transformation parameters. Such a matrix can be constructed as the
second-order polynomial in the world-sheet projections of Cartan
1-forms \cite{AF}
\begin{widetext}
\begin{equation}
K_{ij}{}^\alpha_{\hat a}{}_\beta^{\hat b}=\left(\begin{array}{cc}
\delta^\alpha_\beta(G_{i0'}{}^{m'}G_{j0'm'}\delta^b_a+\Omega_{ia}{}^4\Omega_{j4}{}^b) & iG_{i0'm'}\gamma^{0'm'\alpha}{}_\beta\varepsilon_{acb}\Omega_{j4}{}^c \\
-iG_{i0'm'}\gamma^{0'm'\alpha}{}_\beta\varepsilon^{acb}\Omega_{jc}{}^4
&
\delta^\alpha_\beta(G_{i0'}{}^{m'}G_{j0'm'}\delta_b^a+\Omega_{i4}{}^a\Omega_{jb}{}^4)
\end{array}\right)
\end{equation}
\end{widetext}
and is used in the $\kappa$-symmetry transformation rules for the fermionic 1-forms
\begin{eqnarray}\label{kappaf}
\mathscr F^\alpha_{\{-\}\hat a}(\delta_\kappa)&=&V^{ij}_-V^{kl}_-K_{jl}{}^\alpha_{\hat a}{}_\beta^{\hat b}\varkappa^\beta_{\{-\}\hat bik},\nonumber\\
\mathscr F^\alpha_{\{+\}\hat
a}(\delta_\kappa)&=&V^{ij}_+V^{kl}_+K_{jl}{}^\alpha_{\hat
a}{}_\beta^{\hat b}\widetilde\varkappa^\beta_{\{+\}\hat bik}.\nonumber\\
\end{eqnarray}
As the bosonic forms are inert under the $\kappa-$symmetry \cite{Witten'86}
\begin{equation}
G_{0'm'}(\delta_\kappa)=0,\qquad\Omega_{4}{}^a(\delta_\kappa)=0,\qquad\Omega_a{}^4(\delta_\kappa)=0,
\end{equation}
the $\kappa-$variation of the action (\ref{action}) obtained by the substitution of (\ref{kappaf}) into (\ref{fermivar})
is compensated by the variation of the auxiliary $2d$ metric
\begin{eqnarray}
\delta_\kappa(\sqrt{-g}g^{ij})&=&2i(\bar{\mathscr F}^{\alpha\hat
a}_{\{-\}k}V^{kl}_-G_{l0'm'}\gamma^{0'm'}_{\alpha\beta}V^{ii'}_+V^{jj'}_+\varkappa^\beta_{\{-\}\hat
ai'j'}\nonumber\\
&+&\bar{\mathscr F}^{\alpha\hat
a}_{\{+\}k}V^{kl}_+G_{l0'm'}\gamma^{0'm'}_{\alpha\beta}V^{ii'}_-V^{jj'}_-\tilde\varkappa^\beta_{\{+\}\hat
ai'j'}).\nonumber\\
\end{eqnarray}
The polynomial structure of $K$ requires the parameters of the $\kappa-$transformation
$\varkappa^\beta_{\{-\}\hat aij}$, $\widetilde\varkappa^\beta_{\{+\}\hat aij}$ to carry the pair of the world-sheet vector indices
instead of one as in the GS case and to satisfy the (anti)selfduality constraints in each index
\begin{equation}
\frac{1}{\sqrt{-g}}g_{ij}V^{jk}_+\varkappa^\beta_{\{-\}\hat akl}=\frac{1}{\sqrt{-g}}g_{lj}V^{jk}_+\varkappa^\beta_{\{-\}\hat aik}=\varkappa^\beta_{\{-\}\hat ail},
\end{equation}
and
\begin{equation}
\frac{1}{\sqrt{-g}}g_{ij}V^{jk}_-\widetilde\varkappa^\beta_{\{+\}\hat akl}=\frac{1}{\sqrt{-g}}g_{lj}V^{jk}_-\widetilde\varkappa^\beta_{\{+\}\hat aik}=\widetilde\varkappa^\beta_{\{+\}\hat ail}.
\end{equation}

On the constraint shell defined by the Virasoro constraints
(\ref{vir}) the rank of the $\kappa-$transformations equals 8 so
that only $\frac13$ parameters act nontrivially. Note that
in the $\kappa-$symmetry transformation rules (\ref{kappaf}) the $2d$
vector indices of the matrix $K$ are contracted with the
world-sheet projectors $V^{ij}_{\pm}$ so only one independent
component of 4 remains. Thus to find the rank of
$K_{\pm\tau\pm\tau}{}^\alpha_{\hat a}{}_\beta^{\hat b}$ one can
solve the eigenvalue problem that amounts to computing the
determinant of $K-\lambda I$ using its block structure
\begin{eqnarray}
\det(K-\lambda I)&=&\det\left(
\begin{array}{cc}
A^\alpha_a{}^b_\beta & B^\alpha_a{}_{\beta b}\\
C^{\alpha a}{}^b_\beta & D^{\alpha a}{}_{\beta b}
\end{array}\right)\nonumber\\
&=&\det A\det (D-CA^{-1}B),\nonumber\\
\end{eqnarray}
where
\begin{eqnarray}
&A^\alpha_a{}^b_\beta=\delta^\alpha_\beta A_a{}^b,\quad
A_a{}^b=(G_{\pm\tau\pm\tau}-\lambda)\delta^b_a+\Omega_{\pm\tau
a}{}^4\Omega_{\pm\tau4}{}^b,&\nonumber\\
&D^{\alpha a}{}_{\beta b}=\delta^\alpha_\beta((G_{\pm\tau\pm\tau}-\lambda)\delta_b^a+\Omega_{\pm\tau4}{}^a
\Omega_{\pm\tau b}{}^4),&\nonumber\\
&B^\alpha_a{}_{\beta b}=iG_{\pm\tau0'm'}\gamma^{0'm'\alpha}{}_\beta\varepsilon_{acb}\Omega_{\pm\tau4}{}^c,&\nonumber\\
&C^{\alpha a}{}^b_\beta=-iG_{\pm\tau0'm'}\gamma^{0'm'\alpha}{}_\beta\varepsilon^{acb}\Omega_{\pm\tau c}{}^4.&\nonumber\\
\end{eqnarray}
The addition of $\lambda I$ renders the matrix $A_a{}^b$ nonsingular,
$\det A=-\lambda(G_{\pm\tau\pm\tau}-\lambda)^2$, and its inverse
is given by
\begin{equation}
A^{-1}{}_b{}^a=\frac{1}{\lambda(G_{\pm\tau\pm\tau}-\lambda)}(\lambda\delta_b^a+\Omega_{\pm\tau
b}{}^4\Omega_{\pm\tau 4}{}^a).
\end{equation}
Then the calculation yields that
\begin{equation}
\det(K-\lambda I)=\lambda^{16}(\lambda-2G_{\pm\tau\pm\tau})^8=0.
\end{equation}
One finds that 8 of 24 eigenvalues of $K$ are nonzero proving that its rank indeed equals 8. So that the matrices $M$ and $K$ are complementary in the
sense that $rankM+rankK=24$.

\section{\label{sec3}Superstring action in the conformal basis}

The introduction of the $(1+2)-$dimensional superconformal group generators (\ref{defconf}), (\ref{defsconf}) (see Appendix B) implies via (\ref{defcartan}) the
introduction of the corresponding 1-forms in the conformal basis
\begin{eqnarray}\label{confcartanb}
&\Delta(d)=G_{0'3}(d),\qquad
\hat\omega_m(d)=-(G_{0'm}(d)+G_{3m}(d)),&\nonumber\\
&\hat c_m(d)=G_{3m}(d)-G_{0'm}(d),\qquad m=0,1,2&\nonumber\\
\end{eqnarray}
and
\begin{equation}\label{confcartanf}
F^\alpha_a(d)={\hat\omega^\mu_a\choose\hat\chi_{\mu a}},\qquad\bar F^{\alpha a}(d)={\bar{\hat\omega}{}^{\mu a}\choose\bar{\hat\chi}{}^a_\mu}.
\end{equation}
So that the expression (\ref{defcartan}) acquires the form
\begin{eqnarray}\label{defcartan'}
\mathscr G^{-1}d\mathscr G&=&
G_{mn}(d)M^{mn}+\hat\omega_m(d)P^m+\hat
c_m(d)K^m\nonumber\\
&+&\Delta(d)D
+\Omega_a{}^b(d)V_b{}^a+\Omega_a{}^4(d)V_4{}^a\nonumber\\
&+&\Omega_4{}^a(d)V_a{}^4 +\Omega_4{}^4(d)V_4{}^4
+\hat\omega^\mu_a(d)Q^a_\mu\nonumber\\
&+&\bar{\hat\omega}{}^{\mu a}(d)\bar Q_{\mu a}+\hat\chi_{\mu
a}(d)S^{\mu a}
+\bar{\hat\chi}{}^a_\mu(d)\bar S^\mu_a.\nonumber\\
\end{eqnarray}
It follows from (\ref{action}) and the definition (\ref{confcartanb}), (\ref{confcartanf}) that the Cartan forms $\hat\omega^m(d)$, $\hat c^m(d)$,
$\Delta(d)$ and $\hat\omega^\mu_a(d)$, $\bar{\hat\omega}{}^{\mu a}(d)$, $\hat\chi_{\mu a}(d)$, $\bar{\hat\chi}{}^a_\mu(d)$ enter the superstring action.
Relevant
MC equations in the conformal basis read
\begin{eqnarray}
d\hat\omega^m&-&2\Delta(d)\wedge\hat\omega^m(d)-2G^m{}_n(d)\wedge\hat\omega^{n}(d)\nonumber\\
&+&2i\hat\omega^\mu_a(d)\wedge\sigma^m_{\mu\nu}\bar{\hat\omega}{}^{\nu a}(d)=0,\nonumber\\
d\hat c^m&+&2\Delta(d)\wedge\hat c^m(d)-2G^m{}_n(d)\wedge\hat c^n(d)\nonumber\\
&+&2i\hat\chi_{\mu a}(d)\wedge\tilde\sigma^{m\mu\nu}\bar{\hat\chi}{}^a_\nu(d)=0,\nonumber\\
d\Delta&-&\hat\omega^m(d)\wedge\hat c_m(d)-i(\bar{\hat\omega}{}^{\mu a}(d)\wedge\hat\chi_{\mu a}(d)\nonumber\\
&+&\hat\omega^\mu_a(d)\wedge\bar{\hat\chi}{}^a_\mu(d))=0,\nonumber\\
d\Omega_a{}^4&+&i\Omega_{+a}{}^b(d)\wedge\Omega_b{}^4(d)-2\varepsilon_{abc}\bar{\hat\omega}{}^{\mu b}(d)\wedge\bar{\hat\chi}{}^c_\mu(d)=0,\nonumber\\
d\Omega_4{}^a&+&i\Omega_4{}^b(d)\wedge\Omega_{+b}{}^a(d)+2\varepsilon^{abc}\hat\omega^\mu_b(d)\wedge\hat\chi_{\mu c}(d)=0,\nonumber\\
\end{eqnarray}
and
\begin{eqnarray}
d\hat\omega^\mu_{\hat a}&-&\Delta(d)\wedge\hat\omega^\mu_{\hat
a}(d)+\frac12\hat\omega^\nu_{\hat a}(d)\wedge
G_{mn}(d)\sigma^{mn}{}_\nu{}^\mu\nonumber\\
&+&\hat\omega^m(d)\wedge\tilde\sigma_{m}^{\mu\nu}\hat\chi_{\nu\hat a}(d)+i{\mathit\Omega}_{\hat a}{}^{\hat b}(d)\wedge\hat\omega^\mu_{\hat b}(d)=0,\nonumber\\
\end{eqnarray}
\begin{eqnarray}
d\hat\chi_{\mu\hat a}&+&\Delta(d)\wedge\hat\chi_{\mu\hat
a}(d)+\frac12G_{mn}(d)\wedge\sigma^{mn}{}_\mu{}^\nu\hat\chi_{\nu\hat
a}(d)\nonumber\\
&-&\hat c_m(d)\wedge\sigma^m_{\mu\nu}\hat\omega^\nu_{\hat
a}(d)+i{\mathit\Omega}_{\hat a}{}^{\hat b}(d)\wedge\hat\chi_{\mu\hat
b}(d)=0,\nonumber\\
\end{eqnarray}
where the fermionic 1-forms have been grouped
\begin{equation}
\hat\omega^\mu_{\hat a}(d)={\hat\omega^\mu_a\choose\bar{\hat\omega}{}^{\mu
a}},\qquad\hat\chi_{\mu\hat a}(d)={\hat\chi_{\mu a}\choose\bar{\hat\chi}{}^a_\mu}
\end{equation}
according to the decomposition of the $SO(6)$ vector representation into the $SU(3)$ irreducible parts.
The elements of the matrix $\mathit\Omega_{\hat a}{}^{\hat b}(d)$ are the components of the $su(4)$ Cartan forms (\ref{su4cf})
\begin{equation}\label{6su4}
\mathit\Omega_{\hat a}{}^{\hat b}(d)=\left(
\begin{array}{cc}
\Omega_a{}^b-\delta_a^b\Omega_c{}^c & \varepsilon_{acb}\Omega_4{}^c\\
-\varepsilon^{acb}\Omega_c{}^4 & -\Omega_b{}^a+\delta_b^a\Omega_c{}^c
\end{array}\right).
\end{equation}
It is antisymmetric w.r.t. the metric
\begin{equation}
H_{\hat a\hat b}=\left(
\begin{array}{cc}
0 & \delta_a^b\\
\delta^a_b & 0
\end{array}\right)
\end{equation}
thus having 15 independent components.

To obtain explicit expressions for the Cartan forms in the conformal basis we consider the following
$OSp(4|6)/(SO(1,3)\times U(3))$ supercoset element \footnote{Similar form of the supercoset element was used in
\cite{PST}, \cite{MTlc}, \cite{Metsaev} to consider the superstring and superbranes on the $AdS\times S$ backgrounds.}
\begin{equation}
\mathscr G=e^{x_mP^m+\theta^\mu_aQ^a_\mu+\bar\theta^{\mu a}\bar Q_{\mu a}}e^{\eta_{\mu a}S^{\mu a}+\bar\eta^a_\mu\bar S^{\mu}_a}e^{z^aV_a{}^4+\bar z_aV_4{}^a}e^{\varphi D}.
\end{equation}
The bosonic real coordinates $x^m$ and $\varphi$ parametrize $AdS_4$,
while 3 complex coordinates $z^a$ and their conjugate $\bar z_a$
parametrize $\mathbb{CP}^3$. The anticommuting coordinates can be
divided into $\theta^\mu_a$, $\bar\theta^{\mu a}$ related to
the Poincare supersymmetry and $\eta_{\mu a}$, $\bar\eta^a_\mu$
related to the conformal supersymmetry. Then the calculation yields
for the Cartan forms associated with the $so(2,3)/so(1,3)$ coset generators
\begin{eqnarray}
\hat\omega^m(d)&=&e^{-2\varphi}\omega^m(d),\nonumber\\
\omega^m(d)&=&dx^m-id\theta^\mu_a\sigma^m_{\mu\nu}\bar\theta^{\nu
a}+i\theta^\mu_a\sigma^m_{\mu\nu}d\bar\theta^{\nu a},\nonumber\\
\end{eqnarray}
\begin{eqnarray}\label{conf3cf}
\hat c^m(d)&=&e^{2\varphi}c^m(d),\nonumber\\
c^m(d)&=&-id\eta_{\mu
a}\tilde\sigma^{m\mu\nu}\bar\eta^a_\nu+i\eta_{\mu
a}\tilde\sigma^{m\mu\nu}d\bar\eta^a_\nu\nonumber\\
&-&2(d\theta_{\mu a}+\frac14\zeta_{\mu
a}(d))\tilde\sigma^{m\mu\nu}\bar\eta^a_\nu(\bar\eta^b_\rho\eta^\rho_b)\nonumber\\
&+&2\eta_{\mu a}\tilde\sigma^{m\mu\nu}(d\bar\theta^a_\nu+\frac14\bar\zeta^a_\nu(d))(\bar\eta^b_\rho\eta^\rho_b),\nonumber\\
\end{eqnarray}
\begin{equation}
\Delta(d)=d\varphi+id\theta^\mu_a\bar\eta^a_\mu+id\bar\theta^{\mu
a}\eta_{\mu a},
\end{equation}
where
\begin{equation}
\zeta^\mu_a(d)=-\tilde\sigma^{m\mu\nu}\omega_m(d)\eta_{\nu
a},\quad\bar\zeta^{\mu
a}(d)=-\tilde\sigma^{m\mu\nu}\omega_m(d)\bar\eta_{\nu}^{a},
\end{equation}
 and for
those associated with the $so(1,3)$ generators
\begin{eqnarray}
G^{mn}&=&-i(d\theta^\mu_a+\frac12\zeta^\mu_a(d))\sigma^{mn}{}_\mu{}^\nu\bar\eta^a_\nu\nonumber\\
&-&i(d\bar\theta^{\mu a}+\frac12\bar\zeta^{\mu
a}(d))\sigma^{mn}{}_\mu{}^\nu\eta_{\nu a}.\nonumber\\
\end{eqnarray}
For the $su(4)$ Cartan form matrix (\ref{6su4}) we find
\begin{equation}
\mathit\Omega_{\hat a}{}^{\hat b}(d)=\mathit\Omega_{\mathbf b\hat a}{}^{\hat b}(d)+\mathit\Omega_{\mathbf f\hat a}{}^{\hat b}(d).
\end{equation}
The bosonic contribution is given by
\begin{eqnarray}
\mathit\Omega_{\mathbf b\hat a}{}^{\hat b}(d)&=&iT_{\hat
a}{}^{\hat c}d\bar T_{\hat c}{}^{\hat b}\nonumber\\
&=&\left(
\begin{array}{cc}
\Omega_{\mathbf ba}{}^b-\delta_a^b\Omega_{\mathbf bc}{}^c & \varepsilon_{acb}\Omega_{\mathbf b4}{}^c\\
-\varepsilon^{acb}\Omega_{\mathbf bc}{}^4 & -\Omega_{\mathbf bb}{}^a+\delta_b^a\Omega_{\mathbf bc}{}^c
\end{array}\right),\nonumber\\
\end{eqnarray}
where the unitary matrix $T$ equals
\begin{widetext}
\begin{equation}
T_{\hat a}{}^{\hat b}=\left(
\begin{array}{cc}
\delta_a^b\cos{|z|}+\bar z_az^b\frac{(1-\cos{|z|})}{|z|^2} & i\varepsilon_{acb}z^c\frac{\sin{|z|}}{|z|}\\
-i\varepsilon^{acb}\bar z_c\frac{\sin{|z|}}{|z|} & \delta^a_b\cos{|z|}+z^a\bar z_b\frac{(1-\cos{|z|})}{|z|^2}
\end{array}\right),\quad |z|^2=z^a\bar z_a.
\end{equation}
So that the explicit form of the entries of
$\mathit\Omega_{\mathbf b\hat a}{}^{\hat b}(d)$
 is given by
\begin{eqnarray}
\Omega_{\mathbf ba}{}^b(d)&=&i\frac{(1-\cos{|z|})}{|z|^2}(\bar z_adz^b-d\bar z_az^b)-i\bar z_az^b\frac{(1-\cos{|z|})^2}{2|z|^4}(dz^c\bar z_c-z^cd\bar z_c),\nonumber\\
\Omega_{\mathbf ba}{}^4(d)&=&d\bar z_a\frac{\sin{|z|}}{|z|}+\bar z_a\frac{\sin{|z|}(1-\cos{|z|})}{2|z|^3}(dz^c\bar z_c-z^cd\bar z_c)+\bar z_a\left(\frac{1}{|z|}-\frac{\sin{|z|}}{|z|^2}\right)d|z|,\nonumber\\
\Omega_{\mathbf
b4}{}^a(d)&=&dz^a\frac{\sin{|z|}}{|z|}+z^a\frac{\sin{|z|}(1-\cos{|z|})}{2|z|^3}(z^cd\bar
z_c-dz^c\bar
z_c)+z^a\left(\frac{1}{|z|}-\frac{\sin{|z|}}{|z|^2}\right)d|z|.\nonumber\\
\end{eqnarray}
\end{widetext}
The fermionic contribution can be presented as
\begin{eqnarray}\label{psi}
\mathit\Omega_{\mathbf f\hat a}{}^{\hat b}(d)&=&(T\Psi(d)\bar
T)_{\hat a}{}^{\hat b},\nonumber\\
\Psi_{\hat a}{}^{\hat b}(d)&=&\left(
\begin{array}{cc}
\Psi_{a}{}^b-\delta_a^b\Psi_{c}{}^c & \varepsilon_{acb}\Psi_{4}{}^c\\
-\varepsilon^{acb}\Psi_{c}{}^4 & -\Psi_{b}{}^a+\delta_b^a\Psi_{c}{}^c
\end{array}\right),\nonumber\\
\end{eqnarray}
where the entries of $\Psi_{\hat a}{}^{\hat b}(d)$ equal
\begin{eqnarray}
\Psi_a{}^b(d)&=&2(d\theta^\mu_a+\frac12\zeta^\mu_a(d))\bar\eta^b_\mu-2(d\bar\theta^{\mu
b}+\frac12\bar\zeta^{\mu b}(d))\eta_{\mu a}\nonumber\\
&-&\delta_a^b((d\theta^\mu_c+\frac12\zeta^\mu_c(d))\bar\eta^c_\mu-(d\bar\theta^{\mu
c}+\frac12\bar\zeta^{\mu c}(d))\eta_{\mu c}),\nonumber\\
\Psi_a{}^4(d)&=&2\varepsilon_{abc}(d\bar\theta^{\mu b}+\frac12\bar\zeta^{\mu b}(d))\bar\eta^c_\mu,\nonumber\\
\Psi_4{}^a(d)&=&-2\varepsilon^{abc}(d\theta^\mu_b+\frac12\zeta^\mu_b(d))\eta_{\mu
c}.\nonumber\\
\end{eqnarray}
The expressions for the fermionic Cartan forms can be brought to the form
\begin{eqnarray}
&\left(\begin{array}{c}\hat\omega^\mu_a\\
\bar{\hat\omega}{}^{\mu a}
\end{array}\right)
=e^{-\varphi}T_{\hat a}{}^{\hat b}
\left(\begin{array}{c}
\omega^\mu_b\\
\bar\omega^{\mu b}
\end{array}\right)
,&\nonumber\\
&\omega^\mu_b(d)=d\theta^\mu_b+\zeta^\mu_b(d),\quad\bar\omega^{\mu
b}(d)=d\bar\theta^{\mu b}+\bar\zeta^{\mu b}(d)&\nonumber\\
\end{eqnarray}
and
\begin{equation}
\left(\begin{array}{c}
\hat\chi_{\mu a}\\
\bar{\hat\chi}{}_{\mu}^{a}
\end{array}\right)
=e^{\varphi}T_{\hat a}{}^{\hat b}\left(
\begin{array}{c}
\chi_{\mu b}\\
\bar\chi^b_\mu,\end{array}\right)
\end{equation}
where
\begin{eqnarray}
\chi_{\mu a}(d)&=&d\eta_{\mu a}+2i\bar\eta^b_\mu d\theta^\nu_b\eta_{\nu a}+2i\eta_{\mu b}d\bar\theta^{\nu b}\eta_{\nu a}\nonumber\\
&+&i(d\theta_{\mu a}+\zeta_{\mu a}(d))(\eta^\nu_b\bar\eta^b_\nu),\nonumber\\
\bar\chi^a_\mu(d)&=&d\bar\eta^a_\mu+2i\eta_{\mu b}d\bar\theta^{\nu b}\bar\eta^a_\nu+
2i\bar\eta^b_\mu d\theta^\nu_b\bar\eta^a_\nu\nonumber\\
&+&i(d\bar\theta^a_\mu+\bar\zeta^a_\mu(d))(\eta^\nu_b\bar\eta^b_\nu).\nonumber\\
\end{eqnarray}

In terms of the Cartan forms in the conformal basis (\ref{confcartanb}), (\ref{confcartanf}) the superstring action (\ref{action}) acquires the form
\begin{eqnarray}\label{action'}
\mathscr S&=&-\frac12\int
d^2\xi\sqrt{-g}g^{ij}\left(\frac14(\hat\omega^m_i+\hat
c^m_i)(\hat\omega_{mj}+\hat
c_{mj})\right.\nonumber\\
&+&\left.\Delta_i\Delta_j+\frac12(\Omega_{ia}{}^4\Omega_{j4}{}^a+\Omega_{ja}{}^4\Omega_{i4}{}^a)\right)\nonumber\\
&-&\frac{1}{2}\varepsilon^{ij}\int
d^2\xi\left(\hat\omega^\mu_{ia}\varepsilon_{\mu\nu}\bar{\hat\omega}{}^{\nu
a}_{j}+\hat\chi_{i\mu
a}\varepsilon^{\mu\nu}\bar{\hat\chi}{}^a_{j\nu}\right).\nonumber\\
\end{eqnarray}
It has a rather complicated structure with the kinetic term
containing contributions up to the 8th power in the fermions and the
WZ term up to the 6th power. Note, however, that similarly to the
$AdS_5\times S^5$ superstring anticommuting coordinates
$\theta^\mu_a$, $\bar\theta^{\mu a}$ related to the Poincare
supersymmetry enter expressions for the Cartan forms utmost
quadratically and the nonlinear fermionic contribution is due to
$\eta_{\mu a}$, $\bar\eta^a_\mu$ related to the conformal
supersymmetry. For the $AdS_5\times S^5$ superstring there have
been proposed the $\kappa-$symmetry gauges that entirely remove
the coordinates $\eta$ so that the action becomes quadratic
\cite{KT98} or quartic in the fermions \cite{Pesando},
\cite{Rahmfeld}, \cite{MTlc}. This seems to be the simplest known
form of the $AdS_5\times S^5$ superstring action. In the case under
consideration it is impossible to gauge away all 12 coordinates
$\eta$ by the 8-parameter $\kappa-$symmetry transformation. Among the
$SO(1,2)$ covariant gauges one can consider the gauge
\begin{equation}
\eta_{\mu a}={\eta_{\mu A}\choose\eta_{\mu 3}},\qquad\eta_{\mu A}=0,
\end{equation}
where the index $A$ corresponds to the fundamental representation of $SU(2)$,
that removes 8 coordinates $\eta$. In this case the following entries of the matrix (\ref{psi}) $\Psi_1{}^2=\Psi_2{}^1=\Psi_4{}^3=\Psi_3{}^4=0$ turn to zero and the kinetic term of the superstring action (\ref{action'}) becomes utmost of the sixth order in the fermions. The gauge
\begin{equation}
\theta^\mu_a={\theta^\mu_A\choose\theta^\mu_3},\qquad\theta^\mu_3=0,\qquad\eta_{\mu 3}=0
\end{equation}
removes an equal number of $\theta$ and $\eta$ coordinates
\footnote{Another $\kappa$-symmetry gauge condition that
eliminates equal number of the $\theta$ and $\eta$ coordinates
have been considered in \cite{Zarembo}.}. In this gauge vanish the
components of the Cartan forms
$\Psi_{1,2}{}^4=\Psi_4{}^{1,2}=\Psi_{1,2}{}^3=\Psi_3{}^{1,2}=0$
and $\omega^\mu_3=0$, $\chi_{\mu3}=0$. More substantial
simplification can be attained, e.g., by considering the
noncovariant condition
\begin{equation}
\eta_{1a}=0
\end{equation}
that partially fixes the $\kappa-$symmetry gauge freedom. In such a case $\hat c^1=0$, while other components of the Cartan forms (\ref{conf3cf}) become quadratic in fermions and also $\chi_{1a}=\bar\chi_1^a=0$ so that the kinetic term of the action (\ref{action'}) contains the fermionic contributions up to the fourth power and the WZ term up to the second power. Then the remaining freedom can be used to turn to zero extra Cartan form components.

\section{Conclusion}

In the present paper we have considered in the framework of the supercoset approach the superstring action on the
$AdS_4\times\mathbb{CP}^3$ background \cite{AF} in the conformal basis for the Cartan
1-forms motivated by the isomorphism between the $osp(4|6)$
superalgebra and $D=3$ $\mathcal N=6$ superconformal algebra.
We have obtained the expressions for the Cartan forms
explicitly covariant under the $D=3$ $\mathcal N=6$ super-Poincare transformations starting
from the $OSp(4|6)/(SO(1,3)\times U(3))$ supercoset representative
parametrized by the coordinates associated with the $D=3$ $\mathcal N=6$
superconformal generators. These results can be used to establish a
more transparent relation to the field theory side of the ABJM
duality \cite{ABJM}.

We have also derived the $SO(1,3)\times SU(3)$ covariant expression
for the matrix $M$ that enters the equations of motion for the fermions
and have shown that in the general case its rank equals 16
implying via the second Noether theorem the 8-parameter
$\kappa-$symmetry of the action. The form of the matrix $M$ can
be found by inspecting the anticommutation relations of the fermionic
generators of $osp(4|6)$ superalgebra. The complementary matrix $K$ that enters
the $\kappa-$symmetry transformation rules is quadratic in the
world-sheet projections of Cartan forms rather than linear as for
the GS superstring and we have proved in the $SO(1,3)\times SU(3)$
covariant way that the rank of $K$ equals 8. These results outline the similarities
and differences of the supercoset formulation for the superstring on
$AdS_4\times\mathbb{CP}^3$ background and the conventional GS one.

It was suggested in \cite{AF} that the $OSp(4|6)/(SO(1,3)\times
U(3))$ supercoset action could be obtained by partial gauge fixing of
the $\kappa-$symmetry in the full superstring action on
$AdS_4\times\mathbb{CP}^3$ background. However, it is interesting
to note that such supercoset action per se may be viewed as belonging to the family of the models of pointlike \cite{balu}-\cite{BLPS} and
extended  \cite{ZU02}-\cite{BAPV} objects in extended superspaces describing the BPS
states preserving exotic \footnote{That is different from the
$\frac12$ fraction of the space-time supersymmetry preserved by
the conventional particle, string and brane models \cite{Polchinski}.} fractions of
the space-time supersymmetry. Here the role of extra superspace
variables complementing the super-Poincare ones is played by the bosonic $z^a$, $\bar
z_a$, $\varphi$ and fermionic $\eta_{\mu a}$,
$\bar\eta^a_\mu$ coordinates.

As the extension of the presented results one can examine the
supercoset action invariance under the full $D=3$ $\mathcal N=6$
superconformal transformations, derive the corresponding Noether
charges and calculate their algebra. It is of interest by fixing
the gauge freedom to seek for the simplest form of the action to
be compared with that for the $AdS_5\times S^5$ superstring. Novel
insights into the structure of the action and the quantization problem
could also be gained by working out the first-order formulation in
analogy with the GS superstring on flat background \cite{BZstring}
and elaborating on the twistor transform \cite{U}. We hope to
address these issues in future.

\begin{acknowledgments}

The author is obliged to A.A.~Zheltukhin for valuable discussions.

\end{acknowledgments}

\appendix
\section{Spinors and $\gamma-$matrices}

The $D=2+3$ spinor indices are raised and lowered by means of the antisymmetric charge conjugation matrix and its inverse
\begin{equation}
\psi^\alpha=C^{\alpha\beta}\psi_\beta,\qquad\psi_\alpha=C_{\alpha\beta}\psi^\beta.
\end{equation}
The spinor $\psi^\alpha$ is composed of a pair of the $(1+2)$-dimensional spinors
\begin{equation}
\psi^\alpha={\phi^\mu\choose\varphi_\nu},\qquad\psi_\alpha={\varphi_\mu\choose-\phi^\nu}
\end{equation}
and the charge conjugation matrix and its inverse admit the representation in terms of the $2\times2$ unit matrix
\begin{equation}
C_{\alpha\beta}=\left(\begin{array}{cc}
0&\delta_\mu^\nu\\
-\delta^\mu_\nu &0
\end{array}\right),\qquad
C^{\alpha\beta}=\left(\begin{array}{cc}
0&-\delta^\mu_\nu\\
\delta^\nu_\mu &0
\end{array}\right).
\end{equation}
The position of indices of the 2-component spinors can be changed as follows:
\begin{eqnarray}
&\phi^\mu=\varepsilon^{\mu\nu}\phi_\nu,\qquad\varphi_\mu=\varepsilon_{\mu\nu}\varphi^\nu,&\nonumber\\
&\varepsilon_{\mu\nu}\varepsilon^{\nu\lambda}=\delta_\mu^\lambda,\qquad\varepsilon^{12}=\varepsilon_{21}=1.&\nonumber\\
\end{eqnarray}

The Majorana condition in $1+2$ dimensions
\begin{equation}
(\varphi^\mu)^{\dagger}\sigma^{0\mu}{}_\nu=\varepsilon_{\nu\mu}\varphi^\mu
\end{equation}
amounts to the reality of the spinor components in the chosen basis, where $\tilde\sigma^{0\mu\nu}=\delta^{\mu\nu}$. Accordingly in $2+3$ dimensions the Majorana condition
\begin{equation}
(\psi^\alpha)^{\dagger}(\tilde\gamma^{0'}\gamma^0)^\alpha{}_\beta=C_{\beta\alpha}\psi^\alpha
\end{equation}
is satisfied for the spinors composed of a pair of the $(1+2)-$dimensional
Majorana spinors. Because of the relation
$(\tilde\gamma^{0'}\gamma^0)^\alpha{}_\beta
C^{\beta\gamma}=-\delta^{\alpha\gamma}$ it also amounts to the
component by component reality of a spinor.

The $(2+3)-$dimensional $\gamma-$matrices in the Majorana representation can be realized in terms of the $(1+2)-$dimensional real $\gamma-$matrices
\begin{eqnarray}\label{5to3}
&\gamma^{0'}_{\alpha\beta}=-\left(\begin{array}{cc}
\varepsilon_{\mu\nu}&0\\
0&\varepsilon^{\mu\nu}
\end{array}\right),\
\tilde\gamma^{0'\alpha\beta}=-\left(\begin{array}{cc}
\varepsilon^{\mu\nu}&0\\
0&\varepsilon_{\mu\nu}
\end{array}\right),&\nonumber\\
&\gamma^{m}_{\alpha\beta}=\left(\begin{array}{cc}
0&\sigma^m{}_\mu{}^\nu\\
-\sigma^{m\mu}{}_\nu &0
\end{array}\right),&\nonumber\\
&\tilde\gamma^{m\alpha\beta}=\left(\begin{array}{cc}
0&\sigma^{m\mu}{}_\nu\\
-\sigma^{m}{}_{\mu}{}^\nu &0
\end{array}\right),&\nonumber\\
&\gamma^{3}_{\alpha\beta}=\left(\begin{array}{cc}
\varepsilon_{\mu\nu}&0\\
0&-\varepsilon^{\mu\nu}
\end{array}\right),\
\tilde\gamma^{3\alpha\beta}=\left(\begin{array}{cc}
-\varepsilon^{\mu\nu}&0\\
0&\varepsilon_{\mu\nu}
\end{array}\right),&\nonumber\\
\end{eqnarray}
where
\begin{equation}
\sigma^m_{\mu\nu}=(I,\sigma^1,-\sigma^3),\quad\tilde\sigma^{m\mu\nu}=\varepsilon^{\mu\lambda}\varepsilon^{\nu\rho}\sigma^m_{\lambda\rho}=(I,-\sigma^1,\sigma^3).
\end{equation}
They satisfy the Clifford algebra relations
\begin{eqnarray}
&\gamma^{\underline{m}}_{\alpha\beta}\tilde\gamma^{\underline{n}\beta\gamma}
+\gamma^{\underline{n}}_{\alpha\beta}\tilde\gamma^{\underline{m}\beta\gamma}=-2\eta^{\underline{mn}}\delta^\gamma_\alpha,
&\nonumber\\
&\eta^{\underline{mn}}=(-,-,+,+,+),\qquad\tilde\gamma^{\underline{m}\alpha\beta}=C^{\alpha\gamma}C^{\beta\delta}\gamma^{\underline{m}}_{\gamma\delta}&\nonumber\\
\end{eqnarray}
as a result of the $D=1+2$ relations
\begin{equation}
\sigma^m_{\mu\nu}\tilde\sigma^{n\nu\lambda}+\sigma^n_{\mu\nu}\tilde\sigma^{m\nu\lambda}=-2\eta^{mn}\delta_\mu^\lambda.
\end{equation}

The $so(2,3)$ generators are defined as
\begin{equation}
\gamma^{\underline{mn}}{}_\alpha{}^\beta=\frac12(\gamma^{\underline{m}}_{\alpha\gamma}\tilde\gamma^{\underline{n}\gamma\beta}-\gamma^{\underline{n}}_{\alpha\gamma}\tilde\gamma^{\underline{m}\gamma\beta}).
\end{equation}
Their explicit form in terms of the
above introduced $\gamma-$matrices is found to be
\begin{eqnarray}\label{5to3'}
&\gamma^{mn}{}_\alpha{}^\beta=\left(\begin{array}{cc}
\sigma^{mn}{}_\mu{}^\nu &0 \\
0&\tilde\sigma^{mn\mu}{}_\nu
\end{array}\right),&\nonumber\\
&\gamma^{0'm}{}_\alpha{}^\beta=\left(\begin{array}{cc}
0&-\sigma^m_{\mu\nu} \\
\tilde\sigma^{m\mu\nu}&0
\end{array}\right),&\nonumber\\
&\gamma^{3m}{}_\alpha{}^\beta=\left(\begin{array}{cc}
0&\sigma^m_{\mu\nu} \\
\tilde\sigma^{m\mu\nu}&0
\end{array}\right),\
\gamma^{0'3}{}_\alpha{}^\beta=\left(\begin{array}{cc}
\delta^\nu_\mu &0 \\
0&-\delta^\mu_\nu
\end{array}\right),&\nonumber\\
&\sigma^{mn}{}_\mu{}^\nu=\frac12(\sigma^m_{\mu\lambda}\tilde\sigma^{n\lambda\nu}
-\sigma^n_{\mu\lambda}\tilde\sigma^{m\lambda\nu}),\ \tilde\sigma^{mn\mu}{}_\nu=-\sigma^{mn}{}_\nu{}^\mu.&\nonumber\\
\end{eqnarray}

The $(1+3)-$dimensional charge conjugation matrix that enters the
WZ term and its inverse can be realized as
\begin{eqnarray}
&C'_{\alpha\beta}=-i\gamma^{0'}_{\alpha\beta}=i\left(\begin{array}{cc}
\varepsilon_{\mu\nu}&0\\
0&\varepsilon^{\mu\nu}
\end{array}\right),&\nonumber\\
&C'^{\alpha\beta}=i\tilde\gamma^{0'\alpha\beta}=-i\left(\begin{array}{cc}
\varepsilon^{\mu\nu}&0\\
0&\varepsilon_{\mu\nu}
\end{array}\right).&\nonumber\\
\end{eqnarray}
$\Gamma-$matrices in $D=1+3$ dimensions are defined as
\begin{eqnarray}
&\Gamma^{m'}{}_\alpha{}^\beta=-\gamma^{m'}_{\alpha\gamma}C'^{\gamma\beta}:&\nonumber\\
&\Gamma^m{}_\alpha{}^\beta= \left(\begin{array}{cc}
0&-i\sigma^m_{\mu\nu}\\
i\tilde\sigma^{m\mu\nu}&0
\end{array}\right),\ \Gamma^3{}_\alpha{}^\beta=\left(\begin{array}{cc}
i\delta^\nu_\mu&0\\
0&-i\delta^\mu_\nu
\end{array}\right).&\nonumber\\
\end{eqnarray}
They obey the Clifford algebra relations
\begin{equation}
\Gamma^{m'}{}_\alpha{}^\gamma\Gamma^{n'}{}_\gamma{}^\beta+\Gamma^{n'}{}_\alpha{}^\gamma\Gamma^{m'}{}_\gamma{}^\beta
=-2\eta^{m'n'}\delta_\alpha^\beta.
\end{equation}
The matrix $\Gamma^5=\Gamma^0\Gamma^1\Gamma^2\Gamma^3$  then equals
\begin{equation}
\Gamma^5_\alpha{}^\beta=\left(\begin{array}{cc}
0&\varepsilon_{\mu\nu}\\
-\varepsilon^{\mu\nu}&0
\end{array}\right).
\end{equation}

\section{$osp(4|6)$ superalgebra as $D=3$ $\mathcal N=6$ superconformal algebra}

The (anti)commutation relations of the $osp(4|6)$ superalgebra can
be written in the supermatrix form
\begin{eqnarray}
&[O_{\hat K\hat L},O_{\hat M\hat N}\}=i(G_{\hat L\hat M}O_{\hat
K\hat N}+(-)^{lm}G_{\hat L\hat N}O_{\hat K\hat M}&\nonumber\\
&+(-)^{kl}G_{\hat K\hat M}O_{\hat L\hat N}+(-)^{k(l+m)}G_{\hat
K\hat N}O_{\hat L\hat M}),&\nonumber\\
\end{eqnarray}
where
\begin{equation}
G_{\hat L\hat M}=\left(
\begin{array}{cc}
C_{\alpha\beta}&0\\
0& i\delta_{IJ}
\end{array}\right)
\end{equation}
is the orthosymplectic metric composed of the $D=2+3$ charge
conjugation matrix $C_{\alpha\beta}$ and the unit metric
$\delta_{IJ}$ in the vector representation of $SO(6)$. The supermatrix
$O_{\hat M\hat N}$ has the following block structure
\begin{equation}
O_{\hat M\hat N}=\left(\begin{array}{cc} O_{\alpha\beta}&
O_{\alpha J}\\
 O_{I\beta}& O_{IJ}
\end{array}\right)
\end{equation}
with the blocks obeying the reality
\begin{eqnarray}
O^*_{\alpha\beta}=O_{\alpha\beta},\quad O^*_{\alpha J}=-O_{\alpha
J},\nonumber\\
O^*_{I\beta}=-O_{I\beta},\quad O^*_{IJ}=-O_{IJ}\nonumber\\
\end{eqnarray}
and (anti)symmetry
\begin{eqnarray}
O_{\hat M\hat N}=(-)^{mn}O_{\hat N\hat M}&:&
O_{\alpha\beta}=O_{\beta\alpha},\quad O_{\alpha J}=O_{J\alpha},\nonumber\\
&&O_{IJ}=-O_{JI}
\end{eqnarray}
conditions. The block structure of $O_{\hat M\hat N}$ implies that
the (anti)commutation relations of the $osp(4|6)$ superalgebra can
be divided into 5 groups
\begin{eqnarray}\label{so23}
[O_{\alpha\beta},O_{\gamma\delta}]&=&i(C_{\alpha\gamma}O_{\beta\delta}+C_{\alpha\delta}O_{\beta\gamma}\nonumber\\
&+&C_{\beta\gamma}O_{\alpha\delta}+C_{\beta\delta}O_{\alpha\gamma}),\nonumber\\
\end{eqnarray}
\begin{eqnarray}\label{so6}
[O_{IJ},O_{KL}]&=&\delta_{IK}O_{JL}-\delta_{IL}O_{JK}\nonumber\\
&-&\delta_{JK}O_{IL}+\delta_{JL}O_{IK},\nonumber \\
\end{eqnarray}
\begin{equation}
\label{susy} \{O_{\alpha J}, O_{\gamma L}\}=-\delta_{JL}O_{\alpha\gamma}+iC_{\alpha\gamma}O_{JL},
\end{equation}
\begin{equation}
\label{sso23} [O_{\alpha\beta},O_{\gamma L}]=i(C_{\alpha\gamma}O_{\beta L}+C_{\beta\gamma}O_{\alpha L}),
\end{equation}
\begin{equation}
\label{sso6} [O_{IJ},O_{\gamma L}]=\delta_{IL}O_{\gamma
J}-\delta_{JL}O_{\gamma I}.
\end{equation}

The commutation relations of the first group can be cast into the
$so(2,3)$ algebra relations
\begin{equation}
[M^{\underline{kl}},M^{\underline{mn}}]=\eta^{\underline{kn}}M^{\underline{lm}}-\eta^{\underline{km}}M^{\underline{ln}}
-\eta^{\underline{ln}}M^{\underline{km}}+\eta^{\underline{lm}}M^{\underline{kn}}
\end{equation}
by the transformation
$M^{\underline{kl}}=\frac{i}{4}\gamma^{\underline{kl}\alpha\beta}O_{\alpha\beta}$,
$O_{\alpha\beta}=-\frac{i}{2}\gamma^{\underline{mn}}_{\alpha\beta}M_{\underline{mn}}$.
Separating the generators that carry the second time direction
index one arrives at the $AdS_4$ algebra
\begin{eqnarray}
\left[M^{0'm'},M^{0'n'}\right] &=&M^{m'n'},\nonumber\\
\left[M^{0'k'},M^{m'n'}\right]&=&\eta^{k'm'}M^{0'n'}-\eta^{k'n'}M^{0'm'},\nonumber\\
\left[M^{k'l'},M^{m'n'}\right]&=&\eta^{k'n'}M^{l'm'}-\eta^{k'm'}M^{l'n'}\nonumber\\
&-&\eta^{l'n'}M^{k'm'}+\eta^{l'm'}M^{k'n'}.\nonumber\\
\end{eqnarray}
Introducing the $(1+2)-$dimensional dilatation $D$, momentum $P^m$
and conformal boost $K^m$ generators
\begin{eqnarray}\label{defconf}
&D=2M^{0'3},\quad P^m=-(M^{0'm}+M^{3m}),&\nonumber\\
&K^m=M^{3m}-M^{0'm}&\nonumber\\
\end{eqnarray}
the $AdS_4$ algebra commutation relations transform into the $conf_3$ algebra commutation relations
\begin{eqnarray}
\left[P^m, D\right]&=&-2P^m,\qquad [K^m, D]=2K^m,\nonumber\\
\left[P^m,K^n\right]&=&\eta^{mn}D+2M^{mn},\nonumber\\
\left[P^l,M^{mn}\right]&=&\eta^{lm}P^n-\eta^{ln}P^m,
\nonumber\\
\left[K^l,M^{mn}\right]&=&\eta^{lm}K^n-\eta^{ln}K^m,\nonumber\\
\left[M^{kl},M^{mn}\right]&=&\eta^{kn}M^{lm}-\eta^{km}M^{ln}\nonumber\\
&-&\eta^{ln}M^{km}+\eta^{lm}M^{kn}.\nonumber\\
\end{eqnarray}

By converting the $so(6)$ generators into the $su(4)$ generators
\begin{equation}
V_{\hat i}{}^{\hat j}=-\frac{i}{4}O_{IJ}\rho^{IJ}{}_{\hat
i}{}^{\hat j},
\end{equation}
where $\rho^{IJ}{}_{\hat i}{}^{\hat j}=\frac12(\rho^I_{\hat i\hat k}\tilde\rho^{J\hat k\hat j}-\rho^J_{\hat i\hat k}\tilde\rho^{I\hat k\hat j})$, the commutation relations (\ref{so6}) reduce to
\begin{equation}\label{su4}
[V_{\hat i}{}^{\hat j},V_{\hat k}{}^{\hat l}]=i(\delta^{\hat j}_{\hat k}V_{\hat i}{}^{\hat l}-\delta_{\hat i}^{\hat l}V_{\hat k}{}^{\hat j}).
\end{equation}
The $su(4)$ generators can be split into the $u(3)$ generators $V_a{}^b$ and the $su(4)/u(3)$ coset
generators $V_a{}^4$, $V_4{}^a$
\begin{equation}
V_{\hat i}{}^{\hat j}=\left(
\begin{array}{cc}
V_a{}^b& V_a{}^4\\
V_4{}^b& V_4{}^4
\end{array}\right),\qquad V_4{}^4=-V_a{}^a.
\end{equation}
Then the $su(4)$ algebra commutation relations (\ref{su4}) acquire the form
\begin{eqnarray}
\left[V_a{}^4,V_4{}^b\right]&=&i(V_a{}^b+\delta_a^bV_c{}^c),\quad [V_a{}^4,V_b{}^c]=-i\delta_a^cV_b{}^4,\nonumber\\
\left[V_4{}^a,V_b{}^c\right]&=&i\delta_b^aV_4{}^c,\nonumber\\
\left[V_a{}^b,V_c{}^d\right]&=&i(\delta^b_cV_a{}^d-\delta^d_aV_c{}^b).\nonumber\\
\end{eqnarray}

By contracting the $SO(6)$ vector index $I$ of the $osp(4|6)$ fermionic generators $O_{\alpha I}$ with the
$D=6$ antisymmetric chiral $\gamma-$matrices $\rho^I_{\hat
i\hat j}$ and $\tilde\rho^{I\hat i\hat j}$ that satisfy
\begin{equation}
\rho^I_{\hat i\hat j}\tilde\rho^{J\hat j\hat k}+\rho^J_{\hat i\hat j}\tilde\rho^{I\hat j\hat k}=2\delta^{IJ}\delta^{\hat k}_{\hat i},
\end{equation}
the anticommutator (\ref{susy}) is brought to
the form
\begin{eqnarray}\label{susy'}
\{O_{\alpha\hat i\hat j},O_\beta^{\hat k\hat l}\}&=&i(\delta_{\hat
i}^{\hat l}\delta_{\hat j}^{\hat k}-\delta_{\hat i}^{\hat
k}\delta_{\hat j}^{\hat
l})\gamma^{\underline{mn}}_{\alpha\beta}M_{\underline{mn}}\nonumber\\
&+&2C_{\alpha\beta}(\delta^{\hat k}_{\hat i}V_{\hat j}{}^{\hat
l}-\delta^{\hat k}_{\hat j}V_{\hat i}{}^{\hat l}+\delta^{\hat
l}_{\hat j}V_{\hat i}{}^{\hat k}-\delta^{\hat l}_{\hat i}V_{\hat
j}{}^{\hat k}).\nonumber\\
\end{eqnarray}
Performing the $3+1$ split of the $SU(4)$ indices $\hat i=(a,4)$, $\hat
j=(b,4)$, using the duality relations
\begin{eqnarray}
&O_{ab}=-\varepsilon_{abc}O^{4c},\quad
O^{4a}=-\frac12\varepsilon^{abc}O_{bc},&\nonumber\\
&O^{ab}=-\varepsilon^{abc}O_{4c},\quad
O_{4a}=-\frac12\varepsilon_{abc}O^{bc},\quad\varepsilon_{123}=\varepsilon^{123}=1&\nonumber\\
\end{eqnarray}
that stem from the $SU(4)$ duality relations
\begin{equation}
O_{\hat i\hat j}=\frac12\varepsilon_{\hat i\hat j\hat k\hat
l}O^{\hat k\hat l},\qquad\varepsilon_{\hat i\hat j\hat k\hat
l}=\varepsilon_{abc4},
\end{equation}
introducing the $(1+2)-$dimensional supersymmetry $Q_{\mu}^a$, $\bar Q_{\mu a}$ and superconformal $S^{\mu a}$, $\bar S^\mu_a$ generators
\begin{equation}\label{defsconf}
O_{\alpha}^{4a}={Q^a_\mu\choose S^{\mu a}},\qquad O_{\alpha
4a}={\bar Q_{\mu a}\choose\bar S^{\mu}_a},
\end{equation}
substituting the expressions (\ref{5to3'}) and the definition of the $conf_3$ generators (\ref{defconf})
we are able to bring the relations (\ref{susy'}) to the anticommutation
relations of $D=3$ $\mathcal N=6$ superconformal algebra in the $SU(3)$ notation
\begin{eqnarray}
\{Q^a_\mu,\bar Q_{\nu
b}\}&=&2i\delta^a_b\sigma^m_{\mu\nu}P_m,\quad\{S^{\mu a},\bar
S^{\nu}_b\}=
2i\delta^a_b\tilde\sigma^{m\mu\nu}K_m, \nonumber\\
\{Q^a_\mu,S^{\nu
b}\}&=&2\delta_\mu^\nu\varepsilon^{abc}V_c{}^4,\quad\{\bar Q_{\mu
a},\bar S^\nu_b\}=-2\delta_\mu^\nu\varepsilon_{abc}V_4{}^c, \nonumber\\
\{Q^a_\mu,\bar S^\nu_b\}&=&-i\delta^a_b\delta_\mu^\nu
D+i\delta^a_b\sigma^{mn}{}_\mu{}^\nu
M_{mn}\nonumber\\
&-&2\delta^\nu_\mu(V_b{}^a-\delta^a_bV_c{}^c), \nonumber\\
\{\bar Q_{\mu a},S^{\nu b}\}&=&-i\delta_a^b\delta_\mu^\nu
D+i\delta_a^b\sigma^{mn}{}_\mu{}^\nu
M_{mn}\nonumber\\
&+&2\delta^\nu_\mu(V_a{}^b-\delta_a^bV_c{}^c).\nonumber\\
\end{eqnarray}

The commutators of (\ref{sso23}) and (\ref{sso6}) define the properties of the fermionic generators under the $SO(2,3)$ and $SO(6)$ transformations. In particular, using the definition of $D=3$ $\mathcal N=6$ generators (\ref{defconf}) and (\ref{defsconf}) we get
\begin{eqnarray}
\left[D,Q^a_\mu\right]&=&Q^a_\mu,\quad [D,\bar Q_{\mu a}]=\bar
Q_{\mu
a},\nonumber\\
\left[M^{mn},Q^a_\mu\right]&=&\frac12\sigma^{mn}{}_\mu{}^\nu
Q^a_\nu,\nonumber\\
\left[M^{mn},\bar Q_{\mu
a}\right]&=&\frac12\sigma^{mn}{}_\mu{}^\nu
\bar Q_{\nu a},\nonumber\\
\left[K^m,Q^a_\mu\right]&=&\sigma^m_{\mu\nu}S^{\nu a},\quad [K^m,\bar Q_{\mu a}]=\sigma^m_{\mu\nu}\bar S^{\nu}_a,\nonumber\\
\left[D,S^{\mu a}\right]&=&-S^{\mu a},\quad [D,\bar S^{\mu}_{a}]=-\bar S^{\mu}_{a},\nonumber\\
\left[M^{mn},S^{\mu a}\right]&=&-\frac12S^{\nu
a}\sigma^{mn}{}_\nu{}^\mu,\nonumber\\
\left[M^{mn},\bar S^{\mu}_{a}\right]&=&-\frac12\bar S^\nu_a
\sigma^{mn}{}_\nu{}^\mu,\nonumber\\
\left[P^m,S^{\mu a}\right]&=&-\tilde\sigma^{m\mu\nu}Q^a_\nu,\quad [P^m,\bar S^{\mu a}]=-\tilde\sigma^{m\mu\nu}\bar Q_{\nu a}\nonumber\\
\end{eqnarray}
and
\begin{eqnarray}
\left[V_a{}^b,Q^c_\mu\right]&=&\frac{i}{2}\delta_a^bQ^c_\mu-i\delta^c_aQ^b_\mu,\quad [V_4{}^a,Q^b_\mu]=i\varepsilon^{abc}\bar Q_{\mu c},\nonumber\\
\left[V_a{}^b,\bar Q_{\mu c}\right]&=&-\frac{i}{2}\delta_a^b\bar
Q_{\mu
c}+i\delta^b_c\bar Q_{\mu a},\nonumber\\
\left[V_a{}^4,\bar Q_{\mu
b}\right]&=&-i\varepsilon_{abc}Q^c_\mu,\nonumber\\
\left[V_a{}^b,S^{\mu c}\right]&=&\frac{i}{2}\delta_a^bS^{\mu
c}-i\delta^c_aS^{\mu b},\quad [V_4{}^a,S^{\mu
b}]=i\varepsilon^{abc}\bar S^{\mu}_{c},\nonumber\\
\left[V_a{}^b,\bar S^{\mu}_{c}\right]&=&-\frac{i}{2}\delta_a^b\bar
S^{\mu}_{c}+i\delta^b_c\bar S^{\mu}_{a},\quad [V_a{}^4,\bar
S^{\mu}_{b}]=-i\varepsilon_{abc}S^{\mu c}.\nonumber\\
\end{eqnarray}


\begin{thebibliography}{99}
\bibitem{Maldacena}
J.~M.~Maldacena, Adv.\ Theor.\ Math.\ Phys.\ \textbf{2}, 231 (1998).
\bibitem{GKP98}
S.~S.~Gubser, I.~R.~Klebanov, and A.~M.~Polyakov, Phys.\ Lett.\ B \textbf{428}, 105 (1998).
\bibitem{Witten:1998qj}
E.~Witten, Adv.\ Theor.\ Math.\ Phys.\ \textbf{2}, 253 (1998).


\bibitem{ABJM}
O.~Aharony, O.~Bergman, D.~L.~Jafferis, and J.~Maldacena, J. High Energy Phys. 10 (2008) 091.

\bibitem{BLG}
J.~Bagger and N.~Lambert, Phys. Rev. D \textbf{75}, 045020 (2007);
A.~Gustavsson, Nucl.\ Phys.\ \textbf{B811}, 66 (2009); J.~A.~Bagger and
N.~Lambert, Phys. Rev. D \textbf{77}, 065008 (2008).

\bibitem{MT}
R.~R.~Metsaev and A.~A.~Tseytlin, Nucl.\ Phys.\ \textbf{B533}, 109 (1998).

\bibitem{Kallosh}
R.~Kallosh, J.~Rahmfeld, and A.~Rajaraman, J. High Energy Phys. 09 (1998) 002.

\bibitem{BPR}
I.~Bena, J.~Polchinski, and R.~Roiban, Phys.\ Rev.\ D \textbf{69}, 046002 (2004).

\bibitem{Wadia}
G.~Mandal, N.~V.~Suryanarayana, and S.R.~Wadia, Phys. Lett. B
\textbf{543}, 81 (2002).

\bibitem{Dorey}
N.~Dorey, Classical Quantum Gravity \textbf{25}, 214003 (2008).

\bibitem{FT}
S.~Frolov and A.~A.~Tseytlin, J. High Energy Phys. 06 (2002) 007.

\bibitem{semiclas}
Bin Chen and Jun-Bao Wu, J. High Energy Phys. 09 (2008) 096;
T.~McLoughlin and R.~Roiban, J. High Energy Phys. 12 (2008) 101;
L.~F.~Alday, G.~Arutyunov, and D.~Bykov, J. High Energy Phys. 11
(2008) 089; C.~Krishnan, J. High Energy Phys. 09 (2008) 092;
T.~McLoughlin, R.~Roiban, and A.~A.~Tseytlin, J. High Energy Phys.
11 (2008) 069.


\bibitem{AF}
G.~Arutyunov and S.~Frolov, J. High Energy Phys. 09 (2008) 129.

\bibitem{Stefanski}
B.~J.~Stefanski, Nucl.\ Phys.\ \textbf{B808}, 80 (2009).

\bibitem{PS}
P.~Fre and P.~A.~Grassi, \eprint{arXiv:0807.0044 [hep-th]};
G.~Bonelli, P.~A.~Grassi, and H.~Safaai, J. High Energy Phys. 10
(2008) 085; R.~D'Auria, P.~Fre, P.~A.~Grassi, and M.~Trigiante,
Phys.\ Rev.\ D \textbf{79}, 086001 (2009).

\bibitem{MTlc}
R.~R.~Metsaev and A.~A.~Tseytlin, Phys.\ Rev.\ D \textbf{63}, 046002 (2001);
R.~R.~Metsaev, C.~B.~Thorn, and A.~A.~Tseytlin, Nucl.\ Phys.\ \textbf{B596}, 151 (2001).

\bibitem{Metsaev}
R.~R.~Metsaev, Classical Quantum Gravity
\textbf{18}, 1245 (2001).

\bibitem{BLS}
M.~A.~Bandres, A.~E.~Lipstein, and J.~H.~Schwarz, J. High Energy Phys. 09 (2008) 027.

\bibitem{GSWnew}
J.~Gomis, D.~Sorokin, and L.~Wulff, J. High Energy Phys. 03 (2009)
015.

\bibitem{GS}
M.~B.~Green and J.~H.~Schwarz, Phys.\ Lett.\  B \textbf{136}, 367 (1984); Nucl.\ Phys.\  \textbf{B243}, 285 (1984).

\bibitem{Berkovits}
N.~Berkovits, M.~Bershadsky, T.~Hauer, S.~Zhukov, and B.~Zweibach, Nucl. Phys. \textbf{B567}, 61 (2000).

\bibitem{RS}
R.~Roiban and W.~Siegel, J. High Energy Phys. 11 (2000) 024.

\bibitem{Penrose}
R.~Penrose, J. Math. Phys. \textbf{8}, 345 (1967);
R.~Penrose and M.~A.~H.~MacCallum, Phys. Rep. \textbf{6}, 241 (1973);
R.~Penrose and W.~Rindler, \textit{Spinors and space-time} (Cambridge University Press, Cambridge, 1986), Vol. 2.

\bibitem{Witten'86}
E.~Witten, Nucl. Phys. \textbf{B266}, 245 (1986).

\bibitem{PST}
P.~Pasti, D.~Sorokin, and M.~Tonin, Phys. Lett. B \textbf{447}, 251 (1999).

\bibitem{KT98}
R.~Kallosh and A.~A.~Tseytlin, J. High Energy Phys. 10 (1998) 016.

\bibitem{Pesando}
I.~Pesando, J. High Energy Phys. 11 (1998) 002.

\bibitem{Rahmfeld}
R.~Kallosh and J.~Rahmfeld, Phys. Lett. B \textbf{443}, 143 (1998).

\bibitem{Zarembo}
K.~Zarembo, \eprint{arXiv:0903.1747 [hep-th]}.

\bibitem{balu}
I.~Bandos and J.~Lukierski, Mod. Phys. Lett. A \textbf{14}, 1257
(1999); I.A.~Bandos, J.~Lukierski, and D.~P.~Sorokin, Phys. Rev. D
\textbf{61}, 045002 (2000).

\bibitem{BLMB12}
I.~Bandos and J.~Lukierski, Lect. Notes Phys. \textbf{539}, 195 (2000).

\bibitem{BLPS}
I.~A.~Bandos, J.~Lukierski, C.~Preitschopf, and D.~P.~Sorokin, Phys. Rev. D \textbf{61}, 065009 (2000).

\bibitem{ZU02}
A.~A.~Zheltukhin and D.~V.~Uvarov, J. High Energy Phys. 08 (2002) 008;
Phys. Lett. B \textbf{545}, 183 (2002).

\bibitem{b02}
I.~Bandos, Phys. Lett. B \textbf{558}, 197 (2003).

\bibitem{BAPV}
I.~A.~Bandos, J.~A.~de Azcarraga, M.~Picon, and O.~Varela, Phys. Rev. D \textbf{69}, 085007 (2004).

\bibitem{Polchinski}
J.~Hughes, J.~Liu, and J.~Polchinski, Phys. Lett. B \textbf{180}, 370 (1986);
J.~Hughes and J.~Polchinski, Nucl. Phys. \textbf{B278}, 147 (1986).

\bibitem{BZstring}
I.~A.~Bandos and A.~A.~Zheltukhin, JETP Lett. \textbf{54}, 421 (1991);
Phys. Lett. B \textbf{288}, 77 (1992).

\bibitem{U}
D.~V.~Uvarov, Classical Quantum Gravity \textbf{23}, 2711 (2006);
\textbf{24}, 5383 (2007).
\end{thebibliography}
\end{document}